# The Motion Equations of Cosmology Need Relativity Revision

Mei Xiaochun

( Institute of Theoretical Physics in Fuzhou, China )

**Abstract** The motion equation of standard cosmology, the Friedmann equation, is based on the Einstein's equations of gravitational fields. However, British physicist E. A. Milne pointed in 1943 that the same equation could be deduced simply based on the Newtonian theory of gravity. It means that the Friedmann equation, in stead of the motion equation of relativity, is actually the Newtonian one in essence. The reason is that two simplified conditions are used in the deduction of the Friedmann equation. One is the Robertson-Walker metric and another is static energy momentum tensor. It is proved that in light of the Robertson-Walker metric, the velocity of light emitted by celestial bodies in the expansive universe would satisfy the Galileo's addition rule. So what the R-W metric describes is actually the space-time structure of classical mechanics, not that of relativity mechanics. On the other hand, because there are relative moving velocities between observers and material in the expensive universe actually, we should consider dynamic energy momentum tensor in cosmology. The Friedmann equations need to be revised. We should establish the motion equation of real relativity to describe the cosmic processes with high expansive speeds. The real meaning of constant $\kappa$ in the R-W metric is discussed. It is prove strictly in mathematics that when scalar factor $R(t) \neq$ constant or $\dot{R}(t) \neq 0$, constant $\kappa$ can not represent space curvature. Even when $\kappa = 0$, the R-W metric still describes curved space. In this way, some deductions of current cosmology, such as the values of Hubble constant, dark material and dark energy densities and so on, should be re-estimated. Only in this way, we can get rid of the current puzzle situation of cosmology. The possible form of relativity metric for cosmology is proposed at last.

## 1. The Friedmann equations are deduced based on the Newtonian theory of gravity

The motion equation of standard cosmology is the Friedmann equations which are based on the Einstein's equations of gravitational fields. Because the Einstein's equations are too complex to be solved, two simplified conditions are used, which is the Robertson-Walker metric. If cosmic constant is not considered, by using these two conditions and based on the Einstein's equations of gravity, we can obtain

$$\frac{\ddot{R}}{R} = -\frac{4\pi G}{3}(\rho + 3p) \qquad \frac{\ddot{R}}{R} + \frac{2\dot{R}^2}{R^2} + \frac{2\kappa}{R^2} = 4\pi G(\rho - p) \qquad (1)$$

Here $R(t)$ is the scalar factor, $\kappa$ is the curvature constant, $\rho$ the density of cosmic material and $p$ is the intensity of pressure. By canceling $\ddot{R}$ from two formulas above, we obtain

$$\frac{\dot{R}^2}{R^2} + \frac{\kappa}{R^2} = \frac{8\pi G}{3}\rho \qquad (2)$$

The first formula in (1) and (2) are the Friedmann equation. However, British physicist E. A. Milne proved in 1943 that the same equations could be deduced simply based on the Newtonian theory of gravity [1]. It means that the Friedmann equation is actually the one of Newtonian mechanics in essence, in stead of one of relativity. It can only be suitable for the situation when the expansive speed of the universe is very small comparing with light's speed. But for the situation when the expansive speed is high, just as for the high red-shift of supernovae, it would not be suitable. Let's first repeat the deduction of E. A. Milne for the Friedmann equation, then to discuss the problems caused by using two simplified conditions.

According to the principle of cosmology, the universe can be considered as a sphere with uniform material density. In light of the Newtonian theory of gravity, the force acted on a body which is located at $r$ point is only relative to the mass contained in the sphere with radius $r$, having nothing to do with the mass outside the sphere. Let the mass of a uniform sphere be $M = 4\pi r^3(t)\rho(t)/3$, $\rho(t)$ be the mass density at time $t$. At the direction of sphere radius, the Newtonian equation of gravity is

$$\frac{d^2 r}{dt^2} = -\frac{GM}{r^2} = -\frac{4G\pi\rho}{3} r \qquad (3)$$

We do not consider For the description of uniformly expansive sphere, it is convenient to use the following coordinate with $r = R(t)\bar{r}$, in which $\bar{r}$ has nothing to do with time. In this way, (3) becomes

$$\ddot{R}(t) = -\frac{4G\pi\rho(t)}{3} R(t) \qquad (4)$$

The formula (4) is the same as the first formula of (1) when pressure intensity $p = 0$ (If $p \neq 0$, we should use the Newtonian hydrodynamics formula to replace the formula (3)). Because spherical mass is invariable in the expansive process, we have $\rho(t)R^3(t) = \rho(t_0)R^3(t_0) =$ constant, here $t_0$ represents current time. Let

$$\ddot{R}(t) = \frac{d\dot{R}(t)}{dt} = \frac{d\dot{R}(t)}{dR(t)}\frac{dR(t)}{dt} = \dot{R}(t)\frac{d\dot{R}(t)}{dR(t)} \qquad (5)$$

Substituting (5) into (4) and taking integral, we obtain

$$\frac{\dot{R}^2}{R^2} + \frac{\kappa}{R^2} = \frac{8\pi G}{3}\rho \qquad (6)$$

Suppose that initial time is $t'$, the integral constant $\kappa = -\dot{R}^2(t')$ which is equivalent to the curvature constant in the Friedmann equation. The formula (6) is completely the same as (2), in which integral constant $\kappa$ is equivalent to curvature constant. Cosmic constant has not been considered in (4) and (6). Einstein introduced cosmic constant in order to obtain a static universe at beginning. Later, when the universal expansion was founded, Einstein suggested give up it. In the current theory, we often either take it as zero, or combine it with material density for convenience. So it is obvious that the Friedmanns equation (6) is actually the direct result of the Newtonian theory of gravity, not one of relativity, for no any revised factor of relativity is contained in it. As we known that the Newtonian theory is always considered as the non-relativity approximation of the Einstein's theory. When speed is very small or gravity is very weak, the Einstein's theory degenerates into the Newtonian one.

## 2. The R-W metric leads to the Galileo's addition rule of light's velocity

Then we prove that the R-W metric leads to the Galileo's addition rule of light's velocity. According to the principle of cosmology, our universe is uniform and isotropy. The Robertson-Walker metric is considered with the biggest symmetry with form

$$ds^2 = c^2 dt^2 - R^2(t)\left(\frac{d\bar{r}^2}{1-\kappa\bar{r}^2} + \bar{r}^2 d\theta^2 + \bar{r}^2 \sin^2\theta\, d\varphi^2\right) \tag{7}$$

Let's first discuss the situation with $\kappa = 0$, the metric becomes

$$ds^2 = c^2 dt^2 - R^2(t)\left(d\bar{r}^2 + \bar{r}^2 d\theta^2 + \bar{r}^2 \sin^2\theta\, d\varphi^2\right) \tag{8}$$

In this case, (8) is considered to describe flat space-time. Suppose that the observer is at rest at the original point of reference frame and there is a light source fixed at point $\bar{r}$. When light emitted by light's source moves along radius direction, we have $ds = 0$ and $d\theta = d\varphi = 0$. According to (8), we have

$$\frac{d\bar{r}}{dt} = \pm\frac{c}{R(t)} \tag{9}$$

For flat space, real distance between observer and light's source is $r(t) = R(t)\bar{r}$. For the light's source, $\bar{r}$ does not change with time. But for the light emitted by light's source, $\bar{r}$ changes with time as described in (9). The velocity of light source relative to observer at rest at original point of reference frame is

$$V(t) = \frac{dr(t)}{dt} = \dot{R}(t)\bar{r} \tag{10}$$

Therefore, the velocity of light emitted by light's source relative to observer is

$$V_c(t) = \frac{dr(t)}{dt} = \bar{r}\frac{d}{dt}R(t) + R(t)\frac{d\bar{r}}{dt} = \dot{R}(t)\bar{r} \pm c = V(t) \pm c \tag{11}$$

It means that according to the R-W metric, we lead to the Galileo's addition rule of light's velocity in classical mechanics, i.e., the velocity of light is relative to that of light's source. This result violates all modern physical experiments and astronomic observations, and is completely impossible.

As for the curve space with $\kappa \neq 0$, let $ds = 0$ and $d\theta = d\varphi = 0$ in (7), we have

$$\frac{d\bar{r}}{dt} = \pm\frac{c\sqrt{1-\kappa\bar{r}^2}}{R(t)} \quad \text{or} \quad \frac{d\bar{r}}{\sqrt{1-\kappa\bar{r}^2}} = \pm\frac{cdt}{R(t)} \tag{12}$$

On the other hand, as we known that the coordinate $\bar{r}$ has no meaning of measurement in curved space. What is meaningful is proper distance or proper length. Suppose that an observer is at rest at the original point of coordinate system. According to the W-R metric, the definition of proper distance between observer and light's source is [2]

$$r(t) = R(t)\int_0^{\bar{r}} \frac{d\bar{r}_1}{\sqrt{1-\kappa\bar{r}_1^2}} = R(t)l(\bar{r}) \qquad l(\bar{r}) = \int_0^{\bar{r}} \frac{d\bar{r}_1}{\sqrt{1-\kappa\bar{r}_1^2}} \tag{13}$$

In which $l(\bar{r})$ corresponds to $\bar{r}$ in the flat universe. For illuminant material moving in the expansive universe, $l(\bar{r})$ also does not change with time. The velocity of illumine material relative to observer is $V(t) = \dot{R}(t)l(\bar{r})$. The velocity of light emitted by illumine material moves in curving expansive space is

$$V_c(t) = \frac{dr(t)}{dt} = \dot{R}(t)\int_0^{\bar{r}} \frac{d\bar{r}_1}{\sqrt{1-\kappa\bar{r}_1^2}} + R(t)\left(\frac{d}{d\bar{r}}\int_0^{\bar{r}} \frac{d\bar{r}_1}{\sqrt{1-\kappa\bar{r}_1^2}}\right)\frac{d\bar{r}}{dt}$$

$$= \dot{R}(t)l(\bar{r}) + \frac{R(t)}{\sqrt{1-\kappa\bar{r}^2}}\frac{d\bar{r}}{dt} = V(t) \pm c \tag{14}$$

So light's velocity still obeys the Galileo's addition rule when it moves in curved space.

However, as we know that the watershed between classical physics and modern physics is just on the invariance principle of light's speed. **The essence of invariance principle of light's speed is that light's speed has nothing to do with the speed of light's source.** Because the R-W metric violates this principle, it can not be used as the space-time frame for modern cosmology which is considered as the theory of relativity. It can only be considered as the metric of classical mechanics. Especially when the expansive speed of universe is very high, a great error would be caused. In fact, in the current cosmology, we always consider light's speed as a constant when we calculate concrete problems. If light's velocity satisfies the Galileo's addition rule, many problems would be re-calculated. The situations would become very serious! **Even though in strong gravitational fields, perhaps light's speed may be less or exceed its speed in vacuum, it can only be relative to the strength of gravitational field, still having nothing to do with the speed of light's source!** It can yet not satisfy the Galileo's addition rule in this case. This is the most foundational rule of modern physics! If this principle is violated, the theory would not be considered as relativity one again.

## 3. The constant curvature problem of the R-W metric

It is seen from the first part that the Friedmann equation of cosmology can be deduced either from the Newtonian theory of gravity or the Einstein's equation of gravity connecting with the R-W metric. However, the Newtonian theory of gravity is based on flat space-time, but the Einstein's theory of gravity is based on curved space-time. Both have a great difference. Based on the Newtonian theory, the constant $\kappa$ in the Friedmann equation is relative to initial speed, but based on the Einstein's theory, $\kappa$ is considered as curvature constant. The different understandings of constant $\kappa$ would lead to different results and cause great effects in cosmology. We discuss this problem below and prove that under general situations with $\dot{R}(t) \neq 0$ or $R(t) \neq$ constant, $\kappa$ does not represent space curvature. That is to say, no matter whether $\kappa$ is equal to zero or not, the R-W metric can only describe curved space-time, not to describe flat space-time. Therefore, if we use the Friedmann equation to describe the expansive universe which is considered to be flat, the equation can only be considered to be one established on the Newtonian theory of gravity based in flat space-time.

According to current understanding, when $\kappa = 0$, (8) is considered to be the flat metric. This is because that the space part of three dimensions shown in the bracket of (8) is flat. By producing a scalar $R(t)$ which is relative to time, the metric would still be flat. However, on the other side, because of $g_{11} \neq$ constant when $R(t) \neq$ constant, (8) can not be flat metric and its curvature can not be zero. What's wrong? As we known hat the curvature has strict definition in mathematics. We should judge flatness of space by strict calculation, not only by apparent phenomena. We write the metric of four dimension space-time as

$$ds^2 = g_{00}dt^2 - R^2(t)\left(d\bar{r}^2 + \bar{r}^2 d\theta^2 + \bar{r}^2 \sin^2\theta d\varphi^2\right) \tag{15}$$

If time $t$ is fixed with $dt = 0$ and $R(t) =$ constant, the formula becomes the metric of flat space-time. If time $t$ is not fixed with $dt \neq 0$ and $R(t) \neq$ constant, but want to use (15) to describe the three dimension space, we can let $g_{00} = 0$. In this case, (15) becomes

$$ds^2 = -R^2(t)\left(d\bar{r}^2 + \bar{r}^2 d\theta^2 + \bar{r}^2 \sin^2\theta d\varphi^2\right) \tag{16}$$

We have

$$g_{00} = 0 \qquad g_{11} = -R^2(t) \qquad g_{22} = -R^2(t)\bar{r}^2 \qquad g_{33} = -R^2(t)\bar{r}^2 \sin^2\theta$$

$$g^{00} = \infty \qquad g^{11} = -\frac{1}{R^2(t)} \qquad g^{22} = -\frac{1}{R^2(t)\bar{r}^2} \qquad g^{33} = -\frac{1}{R^2(t)\bar{r}^2 \sin^2\theta} \tag{17}$$

$$\Gamma_{11}^0 = R\dot{R}g^{00} \qquad \Gamma_{22}^0 = R\dot{R}\bar{r}^2 g^{00} \qquad \Gamma_{33}^0 = R\dot{R}\bar{r}^2 g^{00}\sin^2\theta \qquad \Gamma_{01}^1 = \Gamma_{02}^2 = \Gamma_{03}^3 = \frac{\dot{R}}{R}$$

$$\Gamma_{22}^1 = -\bar{r} \qquad \Gamma_{33}^1 = -\bar{r}\sin^2\theta \qquad \Gamma_{12}^2 = \Gamma_{13}^3 = \frac{1}{\bar{r}} \qquad \Gamma_{33}^2 = -\sin\theta\cos\theta \qquad \Gamma_{23}^3 = ctg\theta \tag{18}$$

Others $\Gamma_{\alpha\beta}^\rho = 0$. Then calculating the curvature tensors by using following formula

$$R_{\alpha\beta\sigma}{}^\rho = \frac{\partial}{\partial x^\beta}\Gamma_{\sigma\alpha}^\rho - \frac{\partial}{\partial x^\alpha}\Gamma_{\sigma\beta}^\rho + \Gamma_{n\beta}^\rho \Gamma_{\sigma\alpha}^n - \Gamma_{n\alpha}^\rho \Gamma_{\sigma\beta}^n \tag{19}$$

The sufficient and necessary condition of space flatness in the Riemannian geometry is that the curvature tensors become zero everywhere with $R_{\alpha\beta\sigma}{}^\rho = 0$. However, this is impossible for the metric (16). For example, in light of (18) and (19), we have

$$R_{011}{}^0 = -R\ddot{R}g^{00} \qquad R_{022}{}^0 = -R\ddot{R}\bar{r}^2 g^{00} \qquad R_{033}{}^0 = -R\ddot{R}\bar{r}^2 \sin^2\theta\, g^{00}$$

$$R_{100}{}^1 = \frac{\ddot{R}}{R} \qquad R_{122}{}^1 = -\dot{R}^2 \bar{r}^2 g^{00} \qquad R_{133}{}^1 = \dot{R}^2 g^{00} \tag{20}$$

Only when $\dot{R}(t) = 0$, we have $R_{\alpha\beta\sigma}{}^\rho = 0$. So if $\dot{R}(t) \neq 0$, space metric (16) can not be flat. In fact, the flat space metric of three dimensions is

$$d\sigma^2 = dr^2 + r^2 d\theta^2 + r^2 \sin^2\theta d\varphi^2 \tag{21}$$

Let $r = R(t)\bar{r}$ in (21), we have

$$d\sigma^2 = \dot{R}^2(t)dt^2 + 2R(t)\dot{R}(t)\bar{r}dtd\bar{r} + R^2(t)\left(d\bar{r}^2 + \bar{r}^2 d\theta^2 + \bar{r}^2 \sin^2\theta d\varphi^2\right) \tag{22}$$

In light of the principle of Riemannian geometry, if we can find a coordinate transformation to transform a curved space into flat one, the original space would be flat in essence. Only when such transformation does not exist, the original space can be considered curved really. Therefore, the metric (22) is flat but (16) can not be flat, for we can not find a transformation to transform it into the form of (21) when $\dot{R} \neq 0$.

By referring to the situation of spherical surface of two dimensions, we can see this point clearly. The metric (21) describe a flat space of three dimensions with zero curvature. Let $r = $ constant in (21), we have

$$d\sigma_1^2 = r^2\left(d\theta^2 + \sin^2\theta\, d\varphi^2\right) \tag{23}$$

(23) describes a spherical surface of two dimensions with constant curvature $\kappa = 1/r^2 \neq 0$. However, if $r \neq$ constant, the curvature would change with $r$, not a constant again. The position of $r$ in (23) is just as that of $R(t)$ in (16). When $r$ and $R(t)$ are not constants, the metric (16) and (23) have no constant curvatures again. In the Riemannian geometry, the intuitionistic meaning is that when a vector is moved in parallel form along a loop and retuned to the original position, the vector superposes with original vector. If the two vectors can not superpose, the space would be considered curved. Therefore, in light of above calculation, when space expands in the form of (16), a parallel moving vector would not superpose with original vector. We should recognize space curved shown in (16) in this way.

The general form of four dimension flat space-time metric is

$$ds^2 = c^2 dt^2 - \left(dr^2 + r^2 d\theta^2 + r^2 \sin^2\theta\, d\varphi^2\right) \tag{24}$$

By using coordinate transformation $r(t) = R(t)\bar{r}$, we obtain

$$ds^2 = c^2\left[1 - \frac{\dot{R}^2(t)\bar{r}^2}{c^2}\right]dt^2 - 2R(t)\dot{R}(t)\bar{r}d\bar{r}dt - R^2(t)\left(d\bar{r}^2 + \bar{r}^2 d\theta^2 + \bar{r}^2 \sin^2\theta d\varphi^2\right) \tag{25}$$

In appearance, (25) is curved. By the same reason, (25) is still flat. Therefore, (8) can not be flat, because we can not transform it into (24) when $R(t) \neq$ constant. In light of (19), we have $R_{011}^{\ 0} = -R(t)\ddot{R}(t)$ and $R_{122}^{\ 1} = -\dot{R}^2(t)\bar{r}^2$ and so on. Only when $\dot{R} = 0$ or $R(t) =$ constant, we can have $R_{\alpha\beta\sigma}^{\ \rho} = 0$.

This viewpoint is completely different from the current understanding. In view of the significance of this problem in cosmology, we should re-checkup the real meaning of the constant $\kappa$ in the W-R metric. In the process of deducing the R-W metric [2], we proved that the following space metric of three dimensions has a constant curvature $\kappa$

$$d\sigma^2 = \frac{dr^2}{1 - \kappa r^2} + r^2 d\theta^2 + r^2 \sin^2\theta\, d\varphi^2 \tag{26}$$

We can verify this result by calculating the Riemannian formula of constant curvature directly

$$K = \frac{R_{\alpha\beta\sigma\rho}}{g_{\alpha\sigma}g_{\alpha\rho} - g_{\sigma\beta}g_{\alpha\rho}} \tag{27}$$

According to the current method, we fix time $t$ at first to let $r = R(t)\bar{r} = R\bar{r}$ and write (27) as

$$d\sigma^2 = R^2\left(\frac{d\bar{r}^2}{1 - \kappa R^2 \bar{r}^2} + \bar{r}^2 d\theta^2 + \bar{r}^2 \sin^2\theta\, d\varphi^2\right) \tag{28}$$

The space curvature becomes $\kappa' = \kappa R^2$. When $t \neq$ constant or $R(t) \neq$ constant, we enlarge (28) into

$$d\sigma^2 = R^2(t)\left(\frac{d\bar{r}^2}{1 - \kappa'\bar{r}^2} + \bar{r}^2 d\theta^2 + \bar{r}^2 \sin^2\theta\, d\varphi^2\right) \tag{29}$$

Let $\kappa' \to \kappa$ and enlarge (29) into four dimension space-time, we have

$$ds^2 = c^2 dt^2 - R^2(t)\left(\frac{d\bar{r}^2}{1 - \kappa\bar{r}^2} + \bar{r}^2 d\theta^2 + \bar{r}^2 \sin^2\theta\, d\varphi^2\right) \tag{30}$$

In this way, we obtain the R-W metric. It is obvious that the deduction of the W-R metric is not strict one for it contains some analogy and extending. What is verified strictly is only that the metric (26) has a constant curvature $\kappa$. When $R(t) \neq$ constant, we have not proved that the metrics (29) also has a constant curvature $\kappa'$. We have $\kappa' = \kappa R^2(t)$ actually in (29), which is relative to time, not a constant. Speaking simply, when $\dot{R}(t) \neq 0$, the R-W metric has no any constant curvature! By enlarging (26) into four dimensions, we have

$$ds^2 = c^2 dt^2 - \left(\frac{dr^2}{1 - \kappa r^2} + r^2 d\theta^2 + r^2 \sin^2\theta\, d\varphi^2\right) \tag{31}$$

By introducing $r(t) = R(t)\bar{r}$, we obtain

$$ds^2 = c^2\left[1 - \frac{\dot{R}^2(t)\bar{r}^2}{c^2\left(1 - \kappa R^2(t)\bar{r}^2\right)}\right]dt^2 - \frac{2R(t)\dot{R}(t)\bar{r}dtd\bar{r}}{c\left(1 - \kappa R^2(t)\bar{r}^2\right)}$$

$$- R^2(t)\left(\frac{d\bar{r}^2}{1 - \kappa R^2(t)\bar{r}^2} + \bar{r}^2 d\theta^2 + \bar{r}^2 \sin^2\theta d\varphi^2\right) \tag{32}$$

It is different from the R-W metric. Only let $\dot{R}(t)=0$ or $R(t)=$ constant, as well let $\kappa R^2 \to \kappa$, they can be the same.

So in general situations when $\dot{R}(t) \neq 0$, constant $\kappa$ in the W-R metric can not represent curvature. No matter whether $\kappa$ is equal to zero or not, the W-R metric can only describe the metric of curved space-time. It can not be used to describe flat space-time. Of cause, we can also use (30) to describe the expansive universe with $\dot{R}(t) \neq 0$, but the meaning of constant $\kappa$ should be re-considered. As shown before, we can obtain the Friedmann equations directly based on the Newtonian formula of gravity, in which $\kappa$ is an integral constant. This is just physical meaning of constant $\kappa$ in the Friedmann equations. The current cosmology takes a serious mistake in this problem. Even the relativity revision of motion equation is not considered, this result would cause great influence on cosmology. We discuss this problem below.

At first, according to the current observations (WMAP experiments)[3], our universe is nearly flat. If this conclusion is alright, it would be improper for us to use the R-W metric to describe the expending universe, no matter whether we take $\kappa = 0$. On the other hand, as shown before, we can deduce the Friedmann equations directly based on the Newtonian theory which is based on flat space-time. So if we use the Friedmann equations to describe the expensive universe which is flat, we can only consider the Friedmann equations as ones to be established on the Newtonian theory of gravity and flat space-time, not ones based on curved space-time.

Secondly, if we use the Friedmann equations which is based on the Newtonian theory in flat space-time to describe the expensive universe, the constant $\kappa$ should take values among $0 \sim -1$. We have no reason to demand $\kappa = 0$. If $\kappa$ is a big number, it would cause great influence on the Hubble constant, dark material, dark energy and so on. We discuss this problem simply below. Defining the Hubble constant as $H = \dot{R}(t)/R(t)$, we can write the Friedmann equation (2) as

$$\rho(t) = \frac{3\dot{R}^2(t)}{8\pi G R^2(t)} + \frac{3\kappa}{8\pi G R^2(t)} = \frac{3H^2(t)}{8\pi G} + \frac{3\kappa}{8\pi G R^2(t)} \tag{33}$$

By taking current moment $t_0$ in (26), we let $\rho(t_0) = \rho_0$, $H(t_0) = H_0$ and $R(t_0) = R_0$. By defining the critical density of the universal material as $\rho_c = 3H_0^2/(8\pi G)$, at current moment $t_0$, (26) becomes

$$\rho_0 = \rho_c + \frac{3\kappa}{8\pi G R_0^2} \tag{34}$$

The formula is used to estimate the current material density of the universe. According to observation of WMAP[3], our universe is nearly flat. So, if $\kappa$ represent space-time curvature, we have $\kappa = 0$ for the current universe so that the current material density is equal to the critical density. Defining $\Omega = \rho/\rho_c$, we would have $\Omega_0 = \rho_0/\rho_c = 1$ for the current moment for our universe. However, observation shows that for normal material, we only have $\Omega_0 = 0.04$ which is greatly less than 1, so that dark material is needed to fill the universe.

If constant $\kappa$ does not represent space curvature, in light of the result of the Newtonian theory of gravity in flat space-time, we have $\kappa = -\dot{R}^2(t') = -\dot{R}'^2$. In this way, (34) would become

$$\rho_0 = \rho_c - \frac{3\dot{R}'^2}{8\pi G R_0^2} \tag{35}$$

For the proper value of $\dot{R}'$, we may have $\rho_0 \ll \rho_c$. That is to say, if constant $\kappa$ in the Friedmann equations is not curvature, even we do not consider its relativity reversion, it is also unnecessary for us to think that the current material density of the universe is just equal to critical density. At lest, it becomes unnecessary for us to suppose that the quantity of non-baryon dark material is 5 ~ 6 times more than baryon material in the universe. It would also cause significant effect on the estimation of dark energy density 0.7 which is also based on the precondition $\kappa = 0$, combining with the observation of high redshift of supernova [4]. On the other hand, if $\kappa$ is not curvature, in light of the Friedmann equation (2), we have

$$H^2(t) = \frac{8\pi G \rho(t)}{3} - \frac{\kappa}{R^2(t)} \tag{36}$$

In this case, the Hubble constant is not only relative to material density, but also relative to scalar factor $R(t)$. At present time, by considering $\kappa = -\dot{R}'^2$, we have

$$H_0 = \sqrt{\frac{8\pi G \rho_0}{3} - \frac{\kappa}{R_0^2}} = \sqrt{\frac{8\pi G \rho_0}{3} + \frac{\dot{R}'^2}{R_0^2}} > \sqrt{\frac{8\pi G \rho_0}{3}} \tag{37}$$

By a small material density, we would have the same Hubble constant. In fact, as we know at present that the Hubble constant is actually a quit complex concept, not only a pure constant. All of these problems are relative to the real meaning of constant $\kappa$, if is it a curvature?

## 4. Dynamic energy momentum tensor

On the other hand, in cosmology, energy momentum tensor is taken following form of ideal liquid

$$T_{\mu\nu} = (p + \rho)U_\mu U_\nu - p g_{\mu\nu} \tag{38}$$

Here $U_\mu(t)$ is the four dimension velocity. In light of general relativity, we can choose arbitrary reference frame to describe the expansive universe. In the standard cosmology, we take the following reference frames and define energy momentum tensor in the locally static reference frame to let $U_0(t) = 1$ and $U_i(t) = 0$. In this way, we have $T_{00} = \rho$, $T_{0i} = 0$. It means that we take static energy momentum tensor energy in the Einstein's equation of gravity. We only consider static energy of material without considering momentum of material which moves actually in the expensive universe. This kind of approximation is too simple.

In fact, at any certain moment, observers can only be at rest with some material at a certain point of the expansive universe. They can not be at rest with all material in the expansive universe. In principle, we can take the CMB as a rest reference to do observation or measurement, thought we actually take the earth as the original point of static reference frame. So there certainly exist relative velocities between observers and material in the expansive universe. It is impossible for observes to do observation and measurement in the following reference frames which always keep at rest with all material in the expansive universe. In order to coincide with the practical observation, we should consider dynamic energy momentum tensors. For an object located at point $\bar{r}$, its velocities are $V_1(t) = \dot{R}(t)\bar{r}$ and $V_2(t) = V_3(t) = 0$. So we should define the four dimension velocities below ($c = 1$)

$$U_0 = \frac{1}{\sqrt{1 - \dot{R}^2(t)\bar{r}^2}} \qquad U_1 = \frac{\dot{R}(t)\bar{r}}{\sqrt{1 - \dot{R}^2(t)\bar{r}^2}} \qquad U_2 = U_3 = 0 \tag{39}$$

## 5. The possible form of relativity metric for cosmology

In fact, flat space-time is really with biggest symmetry. We can use (25) to describe the dynamic universe. Let's prove that light's speed is still a constant in light of (25). For a clock fixed at following reference frame, we have $d\bar{r} = d\theta = d\varphi = 0$ and (16) becomes

$$ds^2 = c^2\left(1 - \frac{\dot{R}^2(t)\bar{r}^2}{c^2}\right)dt^2 = c^2 d\tau^2 \tag{40}$$

In this way, we obtain the time delay formula of special relativity with

$$d\tau = \sqrt{1 - \frac{V^2}{c^2}}\,dt \qquad\qquad V = \dot{R}(t)\bar{r} \tag{41}$$

For light's motion, we let $ds = 0$ in (25) and get

$$c^2\left[1 - \frac{\dot{R}^2(t)\bar{r}^2}{c^2}\right] - 2R(t)\dot{R}(t)\bar{r}\frac{d\bar{r}}{dt} - R^2(t)\left(\frac{d\bar{r}}{dt}\right)^2 = 0 \tag{42}$$

or

$$\frac{d\bar{r}}{dt} = -\frac{\dot{R}(t)\bar{r} \pm c}{R(t)} \tag{43}$$

Therefore, when light moves along the direction of radius in the expansive universe, its velocity is

$$V_c = \frac{dr}{dt} = \dot{R}(t)\bar{r} + R(t)\frac{d\bar{r}}{dt} = \pm c \tag{44}$$

It is still constant $c$, has nothing to do with the velocity of light's source. Therefore, (25) can be considered as the space-time metric of relativity.

Then let's discuss the motion equation of cosmology based on (25). We have

$$g_{00} = 1 - \dot{R}^2(t)\bar{r}^2 \qquad g_{01} = g_{10} = -R(t)\dot{R}(t)\bar{r} \qquad g_{11} = -R^2(t)$$

$$g_{22} = -R^2(t)\bar{r}^2 \qquad g_{33} = -R^2(t)\bar{r}^2\sin^2\theta \qquad g_{02} = g_{03} = g_{12} = g_{13} = g_{23} = 0 \tag{45}$$

$$g^{00} = 1 \qquad g^{01} = g^{10} = -\frac{\dot{R}(t)\bar{r}}{R(t)} \qquad g^{11} = -\frac{1 - \dot{R}^2(t)\bar{r}^2}{R^2(t)}$$

$$g^{22} = -\frac{1}{R^2(t)\bar{r}^2} \qquad g^{33} = -\frac{1}{R^2(t)\bar{r}^2\sin^2\theta} \qquad g^{02} = g^{03} = g^{12} = g^{13} = g^{23} = 0 \tag{46}$$

Form the formula $\Gamma^\lambda_{\mu\nu} = g^{\lambda\alpha}(\partial g_{\nu\alpha}/\partial x^\mu + \partial g_{\alpha\mu}/\partial x^\nu - \partial g_{\mu\nu}/\partial x^\alpha)/2$, we obtain

$$\Gamma^1_{00} = \frac{\ddot{R}(t)\bar{r}}{R(t)} \qquad \Gamma^1_{01} = \Gamma^2_{02} = \Gamma^3_{03} = \frac{\dot{R}(t)}{R(t)} \qquad \Gamma^1_{33} = -\bar{r}\sin^2\theta$$

$$\Gamma^1_{22} = -\bar{r} \qquad \Gamma^2_{12} = \Gamma^3_{13} = \frac{1}{\bar{r}} \qquad \Gamma^2_{33} = -\sin\theta\cos\theta \qquad \Gamma^3_{23} = \text{ctg}\,\theta \tag{47}$$

Other $\Gamma^\lambda_{\mu\nu}$ is zero. By using the formulas above, for all partial quantities of $R_{\mu\nu}$, we have

$$R_{\mu\nu} = \partial\Gamma^\lambda_{\mu\lambda}/\partial x^\nu - \partial\Gamma^\lambda_{\mu\nu}/\partial x^\lambda + \Gamma^\alpha_{\mu\lambda}\Gamma^\lambda_{\nu\alpha} - \Gamma^\alpha_{\mu\nu}\Gamma^\lambda_{\lambda\alpha} = 0 \tag{48}$$

By considering dynamic energy momentum tensor in the Einstein's equation of gravitational field

$$R_{\mu\nu} + \lambda g_{\mu\nu} = -8\pi G\left(T_{\mu\nu} - \frac{1}{2}T g_{\mu\nu}\right) \tag{49}$$

we obtain the following results for the partial quantities of $R_{00}$, $R_{10}$ and $R_{11}$ ($c=1$):

$$\lambda \left(1-\dot{R}^2\bar{r}^2\right) = -8\pi G\left\{\frac{\rho+p}{1-\dot{R}^2\bar{r}^2} - \frac{1}{2}(\rho-p)\left(1-\dot{R}^2\bar{r}^2\right)\right\} \tag{50}$$

$$\lambda R = 8\pi G\left\{\frac{(\rho+p)}{1-\dot{R}^2\bar{r}^2} + \frac{1}{2}(\rho-p)R\right\} \tag{51}$$

$$\lambda R^2 = 8\pi G\left\{\frac{(\rho+p)\dot{R}^2\bar{r}^2}{1-\dot{R}^2\bar{r}^2} + \frac{1}{2}(\rho-p)R^2\right\} \tag{52}$$

From (50) and (51), we obtain

$$0 = 8\pi G\left\{\frac{(\rho+p)}{1-\dot{R}^2\bar{r}^2}\left[1 + \frac{R(t)}{1-\dot{R}^2\bar{r}^2}\right]\right\} \tag{53}$$

Because $R(t) > 0$ and $\dot{R}(t)\bar{r} = V < 1$ ($c=1$), we get $\rho = p = 0$ from (53). Substituting the result into (52), we get $\lambda = 0$. Therefore, (25) can only describe the empty universe with zero cosmic constant. Because light's speed is invariable in vacuum, this result is quite rational. Because there exists no material, in light of the law of inertial motion, we have $\ddot{R}(t) = 0$ or $\dot{R}(t) = $ constant. That is to say, the universe with zero comic constant can only be both static or expand in a uniform speed in this case.

Because (25) is one of relativity, when gravity exists, in light of the clue shown in (25), we can look for proper metric to describe the space-time construction of the expansive universe. For example, we can take

$$ds^2 = c^2\left[1 - \frac{Q^2(t)\bar{r}^2}{c^2}\right]dt^2 - 2R(t)Q(t)\bar{r}dtd\bar{r} - R^2(t)\left(d\bar{r}^2 + \bar{r}^2 d\theta^2 + \bar{r}^2 \sin^2\theta d\varphi^2\right) \tag{54}$$

When $Q(t) = 0$, (54) becomes the R-W metric with $\kappa = 0$. When $Q(t) = \dot{R}(t)$, it becomes the metric of flat vacuum. In general situations, when $Q(t) \neq \dot{R}(t)$, it would contain the influence of gravity on space-time. By considering (54) and (39) in the Einstein's equation of gravity, we can introduce the revised items of relativity into (1) and (2). In this way, we may establish the relativity motion equation of cosmology with general significance. The detail will be provided later.

Of cause, if the space-time of the expansive universe is flat in light of the result of WMAP experiments, it is still a problem for us to use the motion equation based on curved metric to describe the expansive universe. The R-W metric with $\kappa = 0$ is still a curved one in fact, we are using a curved metric to describe the flat universe at present. The problem still exists to use the metric (54) of relativity to describe the expansive universe for it is also curved when $Q(t) \neq \dot{R}(t)$. This is actually a foundational problem for cosmology we have to face. We remain these problems for later discussions.

## 6. Conclusion

Though the Einstein's equation of gravity is one of relativity, owing to use two simplified conditions, the Friedmann equation becomes one of the Newtonian mechanics in essence. This is an irrefragable fact. When the speed of the universal expansion is low, the Friedmann equation can be effective. But it is unsuitable for the situations when the expansive speed is high, such as the problems of high red-shift of supernovae. On the other hand, when $R(t) \neq$ constant or $\dot{R}(t) \neq 0$, constant $\kappa$ can not represent curvature and the R-W can only describe curved space. If we use the Friedmann equations to describe the expensive universe which is flat, the equations can only be considered ones to be established on the Newtonian theory in flat space-time. It is obvious that many difficulties appearing in the current cosmology originate from the fact that the Friedmann equation is not one of relativity. We deal with modern cosmology based on the classical Newtonian mechanics actually! The Friedmann equations need relativity revision. By means of it, great influences would be caused on cosmology. For example, it is unnecessary for us to think that the current density of the universal material is just equal to critical density. The values of the Hubble constant, dark material and dark energy densities would be re-estimated. In this way, we may get rid of the current puzzle situation of cosmology.